# THE MEASUREMENT OF LATE-PULSES AND AFTER-PULSES IN THE LARGE AREA HAMAMATSU R7081 PHOTOMULTIPLIER WITH IMPROVED QUANTUM-EFFICIENCY PHOTOCATHODE

S. Aiello(1), M. Anghinolfi(2), A. Balbi(2), M. Brunoldi(2), K. Gracheva(3), A. Grimaldi(1), V. Kulikovskiy(2,3), E. Leonora(1), G. Ottonello(2), D. Sciliberto(1), M. Taiuti(2), Y. Yakovenko(3)

*1) INFN - Sezione di Catania, Viale Andrea Doria 6, 95125 Catania, Italy*
*2) Dipartimento di Fisica dell'Università e INFN- Via Dodecaneso 33, 16146 Genova,Italy*
*3) Moscow State University, Leninskie Gori,119991 Moscow*

## Abstract

In recent years, large underwater telescopes have been designed and realized to measure high energy neutrinos from astrophysical objects. Muon tracks produced by the neutrino interaction in the surrounding medium are reconstructed from the arrival time and the number of photo-electrons of the Cherenkov light measured by the Photomultiplier tubes (PMT) array of the detector.
For a correct reconstruction procedure, both the scattering of the light in the water and the late and after pulses produced in the PMTs must be considered.
In this paper we report on this latter effect which has been measured in our laboratory using a laser in the single photoelectron mode (SPE) on a Hamamatsu R7081MOD 10" PMT with a high quantum efficiency photocathode.
The PMT voltage supply was set to provide the 1 photo-electron peak at 10 pC as during normal operation: in this condition we find that the late-pulse contribution is small but not negligible.

## 1. THE SPURIOUS PULSES ON A PMT

The Cherenkov light of the muon track produces pulses on the PMT anode. The pulses are digitized as hits which have arrival time and charge which is proportional to the number of the detected photons. In addition to these ‘main’ pulses also ‘spurious’ pulses may be present due to different physic processes inside the phototube. They can be divided on 4 groups depending on the arrival time (see Fig. 1):

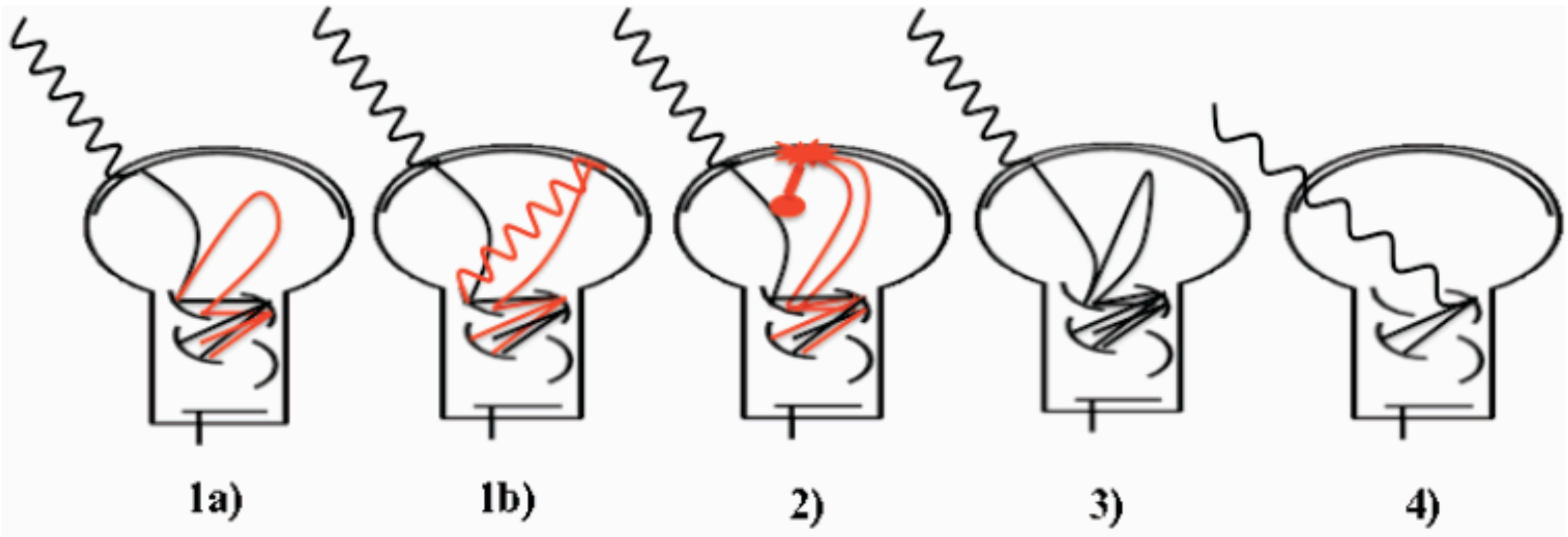

***FIG. 1** – After and late pulses origin on a PMT.*

(1) After-pulses type 1: they appear in the first 10-80ns after the main pulse and can be produced by the luminous reaction on the dynodes while bombarded by the electrons (Fig. 1b) or by the electrons escaped from the dynode in the direction to the photocathode (Fig.. 1a).
(2) After-pulses type 2: appear in the 80ns-16μs interval after the main pulse due to the ionization of the phototube gas by the photoelectron. Different ions give different contributions to the after-pulses time distribution ($H^{+}$, $He^{+}$ and heavy ions $CH^{+4}$ , $O^{+}$, N+2 , $O^{+2}$ , $Ar^{+}$ and $CO^{+2}$ ).
(3) Late-pulses: the primary photoelectron is reflected from the first dynode without a secondary electrons emission, it turns towards the photocathode, makes a loop and only after it creates a cascade of electrons in the dynodes. The arrival time of the hit will be delayed.
(4) Pre-pulses: They appear due to the direct photoelectron emission on the dynodes from the photons which passed the photocathode without interaction. Due to this the main pulse is arriving earlier and with smaller charge.

## 2 THE EXPERIMENTAL SET UP AND DATA ANALYSIS

Measurements were carried out on a R7081MOD 10" PMT with a high quantum efficiency photocathode with a modified active ISEG base.[xxx]. The bare PMT was located in a black light tight box with proper feed through for power supply and anode signal. In the same box an Hamamatsu Picosecond Light Pulser PLP10-044C was located. This laser has a wavelength of 440nm, a pulse width of few ps and a tunable frequency from 1 Hz to 1 MHz. The beam output was connected via an optical fibre to an attenuator and then to a collimator to illuminate the central part of the PMT photocathode as shown in Fig. 2a. The light from the laser was tuned to obtain, at the photocathode window, ~ $10^{-2}$ photons/pulse to produce on average one signal at the PMT anode every 500 light pulses sent by the laser. In these conditions one can assume that the laser is working in the single photoelectron mode. The corresponding signal recorded at the anode of the R7081MOD PMT with the CAEN digitizer as described below is shown in Fig. 2b.

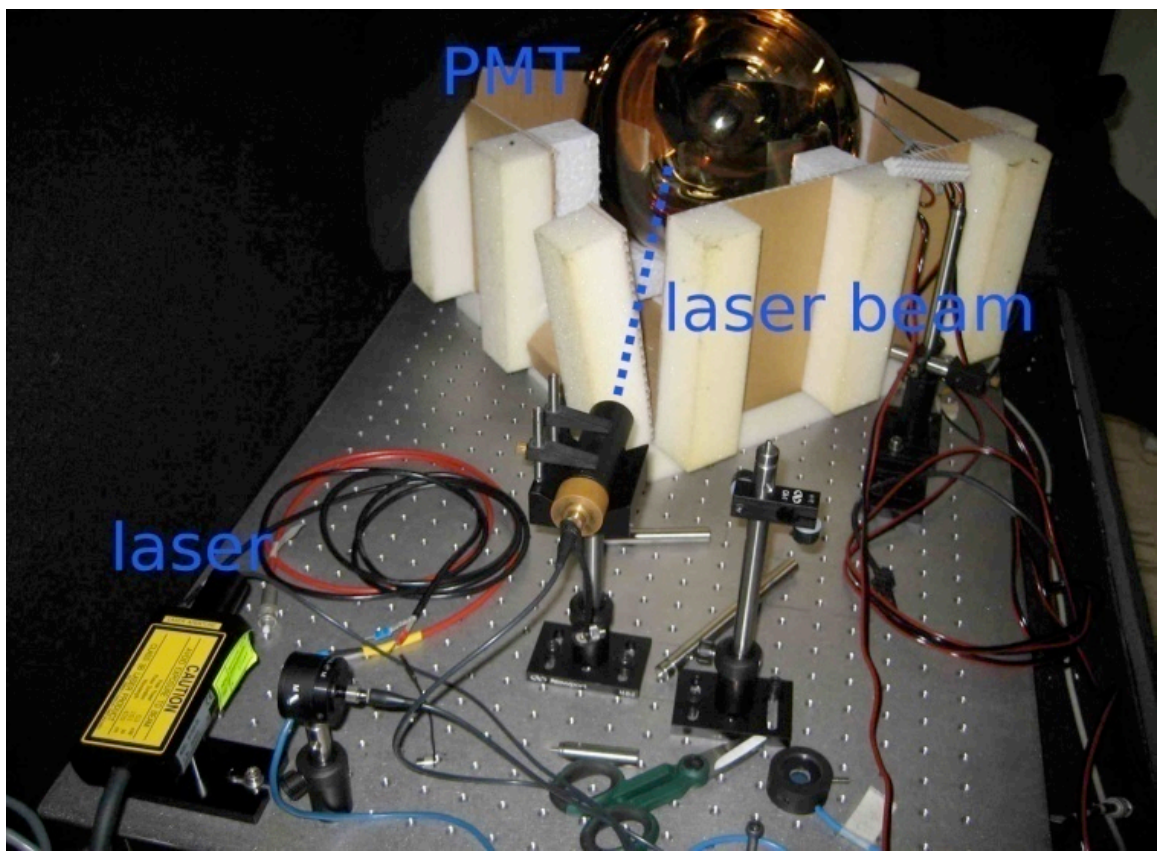

***FIG. 2a** – The experimental set up.*

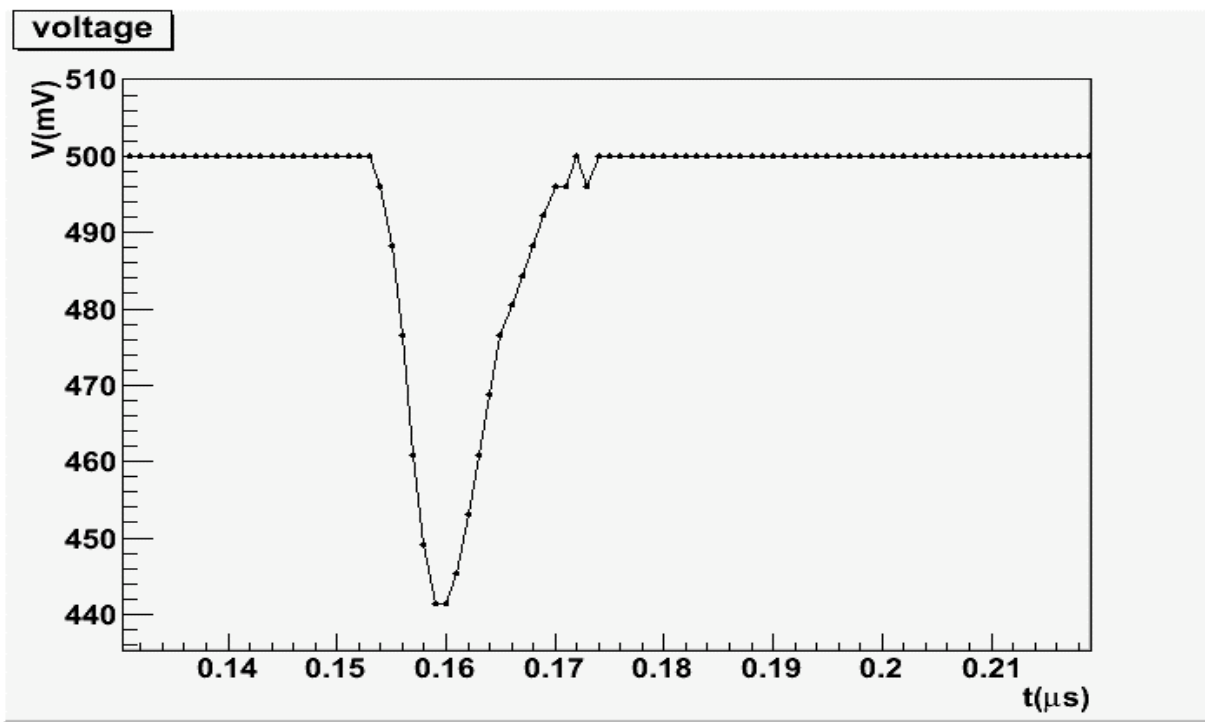

***FIG. 2b** – The anode pulse as measured by the CAEN digitizer.*

The data acquisition system was extremely simple. The output of the PMT anode was split to a 0.3 photo-electron (pe) threshold discriminator and to a fast, 1GHz, CAEN **digitizer V1731**. The output of the discriminator was stretched to 150 ns and put in coincidence with the Laser SYNC out as shown in Fig. 3. In order to exploit the full timing capabilities of the CAEN fast ADC, both the coincidence and the anode signals were sent respectively to channel 0 and 1 of the digitizer. Worth to notice that the logic unit output was always synchronous with respect to the laser SYNC out: the arrival time of the PMT signal was therefore measured relative to the light pulse. In practice, the coincidence NIM signal in channel 0 was used to trigger the acquisition in the following 16 μs time window.

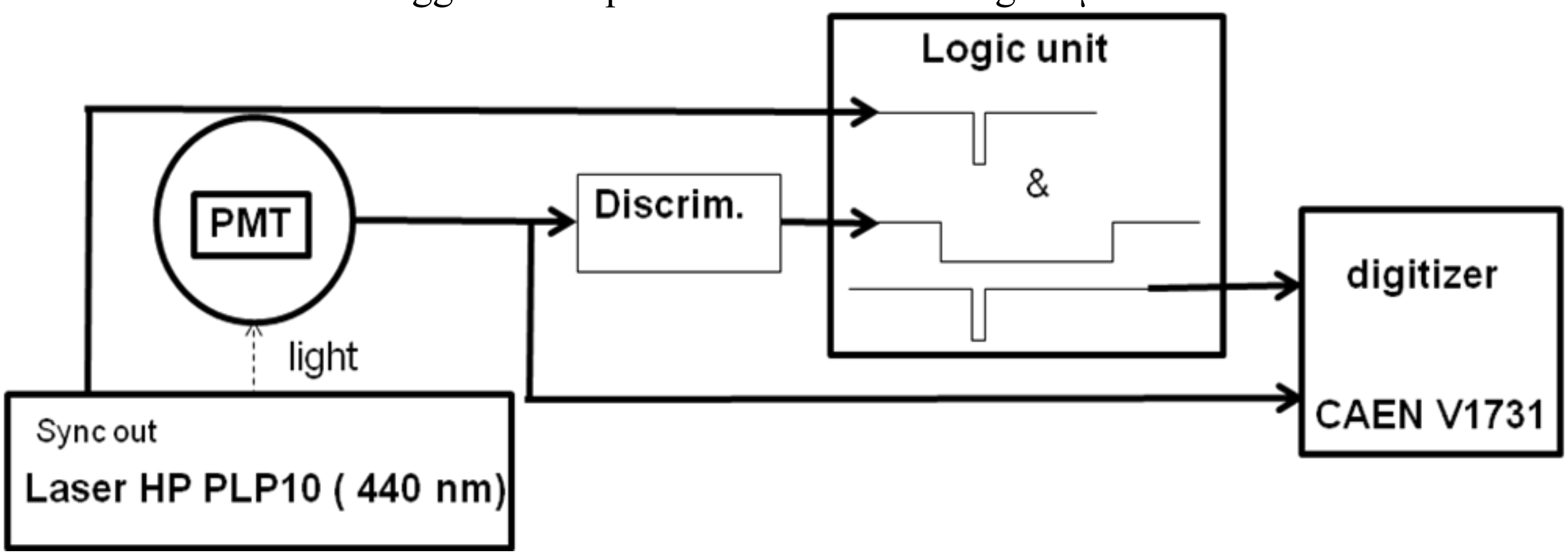

***FIG. 3** – The acquisition system to measure the late and after-pulses.*

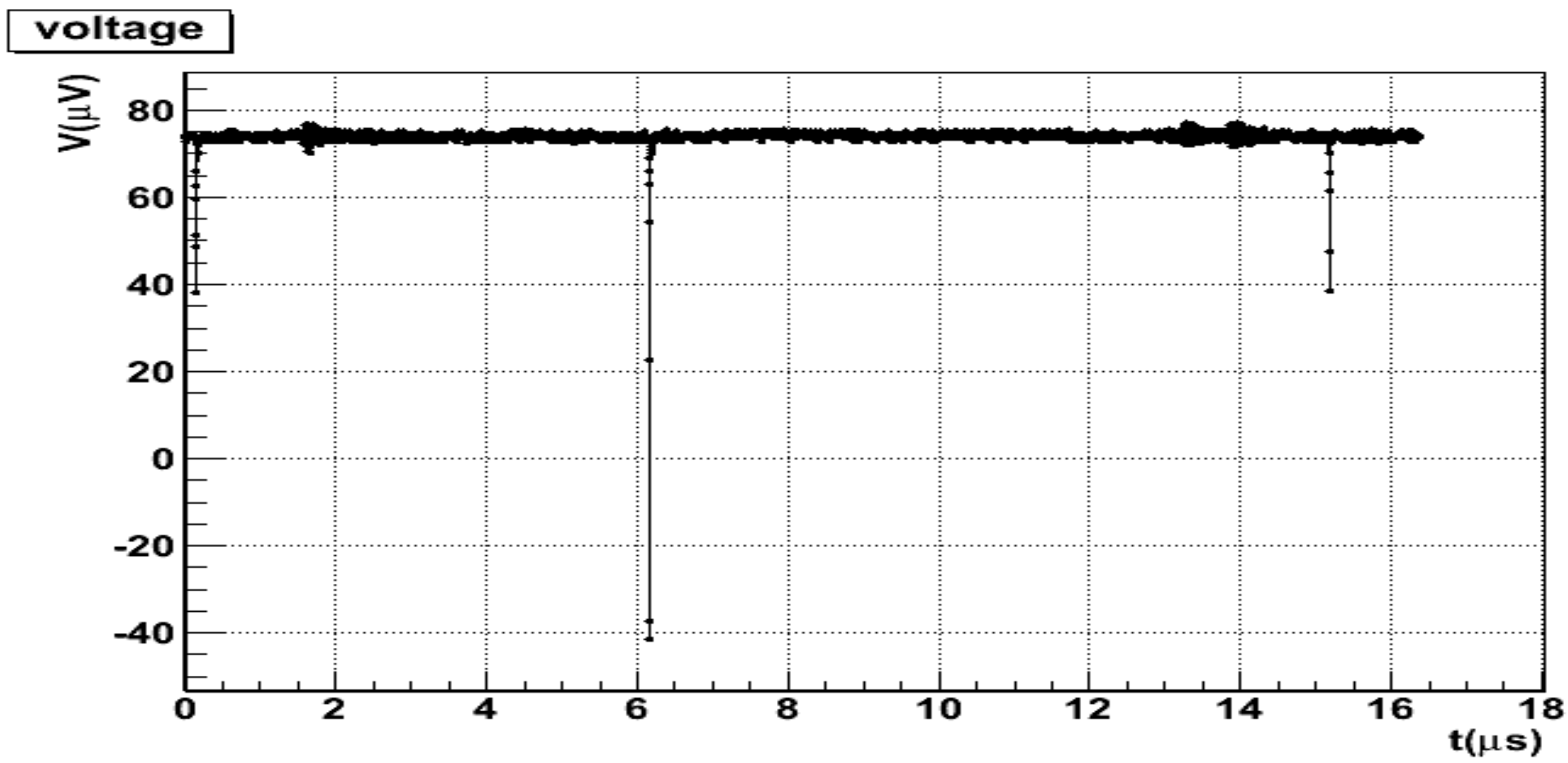

*FIG. 4 – The PMT anode output together with possible after-pulses at 6 and 15 μs from the laser pulse.*

A typical digitized PMT anode output is plotted in Fig. 4 in the 16 μs time interval. Together with the main pulse, after-pulses are seen after ≈6 and 15 μs. Also some electronic noise can be seen at ≈2 μs and ≈13 μs. Sampling rate is 1GHz and voltage is digitized in 0.5mV steps. Charge analysis consists of the next steps:

1. Determination of the pedestal of the signal in the 16 μs interval and subtract it from the signal,
2. Search for all peaks and their charge by iterating time sample by time sample. When the signal is higher than 10mV, corresponding to a threshold of about 0.2-0.3 pe, the program looks for the presence of a peak and finds the left and right boundaries of the peak $T_l$ and $T_r$ where the amplitude of the signal is higher than 10mV. The time of the maximum amplitude is considered as the time of the pulse. Two peaks within a 25 ns time interval are assigned to the same event and the time of the event is the time of the maximum of the two peaks. The charge is evaluated integrating the peak within the Simpson approximation from $T_l$ - 5ns to $T_r$ + 10ns.
3. The hits in first 200ns of the time window are considered as ‘good’ (ie. not after) pulses. The plot of the corresponding charge distribution is shown in Fig. 5. The plot is well fitted with a single gaussian distribution to prove that contributions of 2 or more photo electrons to the main hits are negligible. The left part at low charge is a little bit distorted due to the integration procedure. The mean value of the charge of the 1p.e. peak <q> is found from the fit. After the fit is done, we divide the charge of a hit q to <q> to obtain the charge in pe

Knowing the charge and time of the main pulse and after-pulses we performed several analysis.

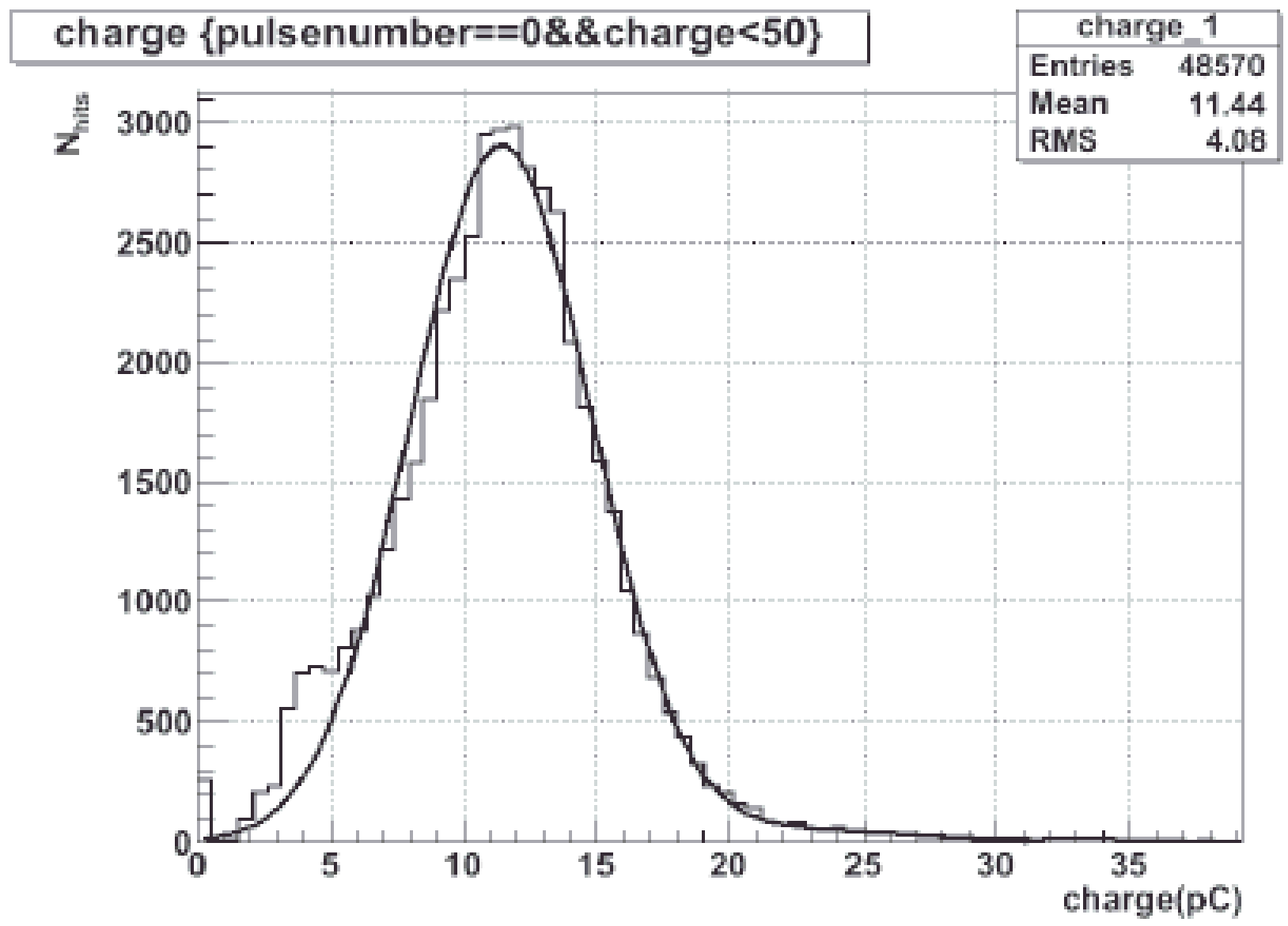

***FIG. 5*** *– The charge distribution of the main pulses: continuous curve is the gaussian fit.*

## 3 LATE AND AFTER-PULSE SEARCH

### 3.1 late-pulses

In Fig. 6a the arrival time of the first hit after the laser pulse is shown in log scale. The peak of the normal pulses from the laser photons is clearly evident at t ~ 100 ns and its FWHM correspond to ~ 2 ns, a bit lower than the expected TTS of 3 ns due to the small spot of the photon laser beam. Random background forms a plateau at the level of 1% in agreement to the observed background rate of 3.4kHz in the absence of the laser signal. The contribution of the late-pulses as deduced from the plot is of the order of 30% when considering arrival times larger than 2σ (~ 2 ns) from the main peak or, more realistically, of about 5% from the minimum of the distribution at 30ns. The origin of the structure at ~ 60 ns from the main peak is not clear. Assuming an uniform electric field of ~ 60 V/cm in the 14 cm distance of the first dynode from the photocathode, the 60 ns roughly corresponds to the 'round trip' of one electron from the first dynode to the photocathode. On the left side from the peak, the pre-pulses contribution is not seen though statistics may be insufficient. In Fig. 6b the time of the late pulse arrival is plotted together with its amplitude. No evident correlation between these two quantities is seen in the plot, the average charge of the late pulses being the same as the main pulses.

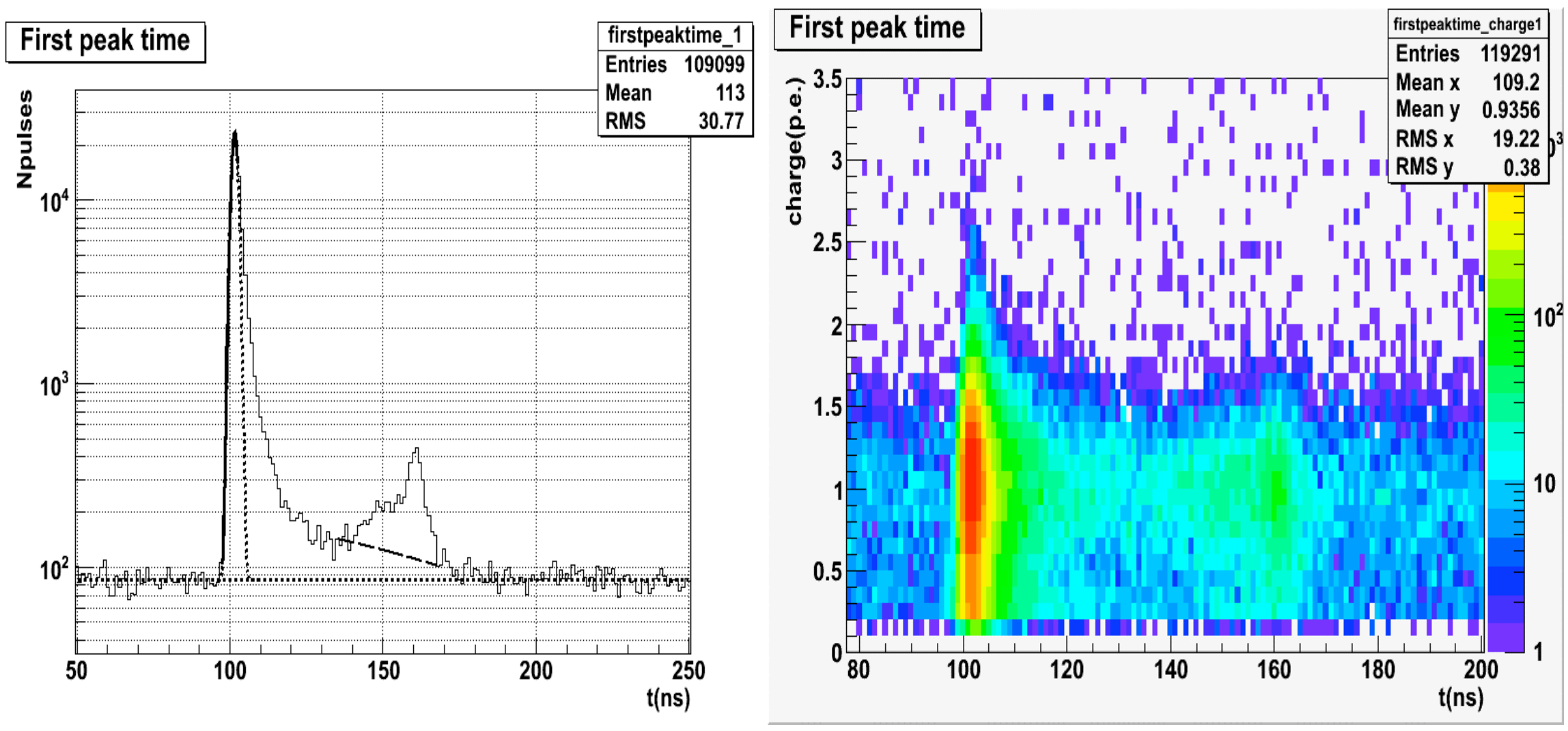

***FIG. 6*** *– a) late pulse distribution in our improved quantum efficiency phototube; b) the amplitude of the late pulse distribution.*

### 3.2. After-pulses search

For the search of after-pulses we took only events with the main hit in the time window from 90ns to 120ns from beginning of the frame to exclude late pulses events (Fig. 6). A threshold of 0.3 pe was also applied on pulses and after-pulses.

In Fig. 7 the time vs. amplitude distribution of the after-pulses is reported. The left plot is limited to the analysis of after-pulses occurring up to 200 ns from the laser pulse i.e. of the type 1a and 1b of Fig. 1.

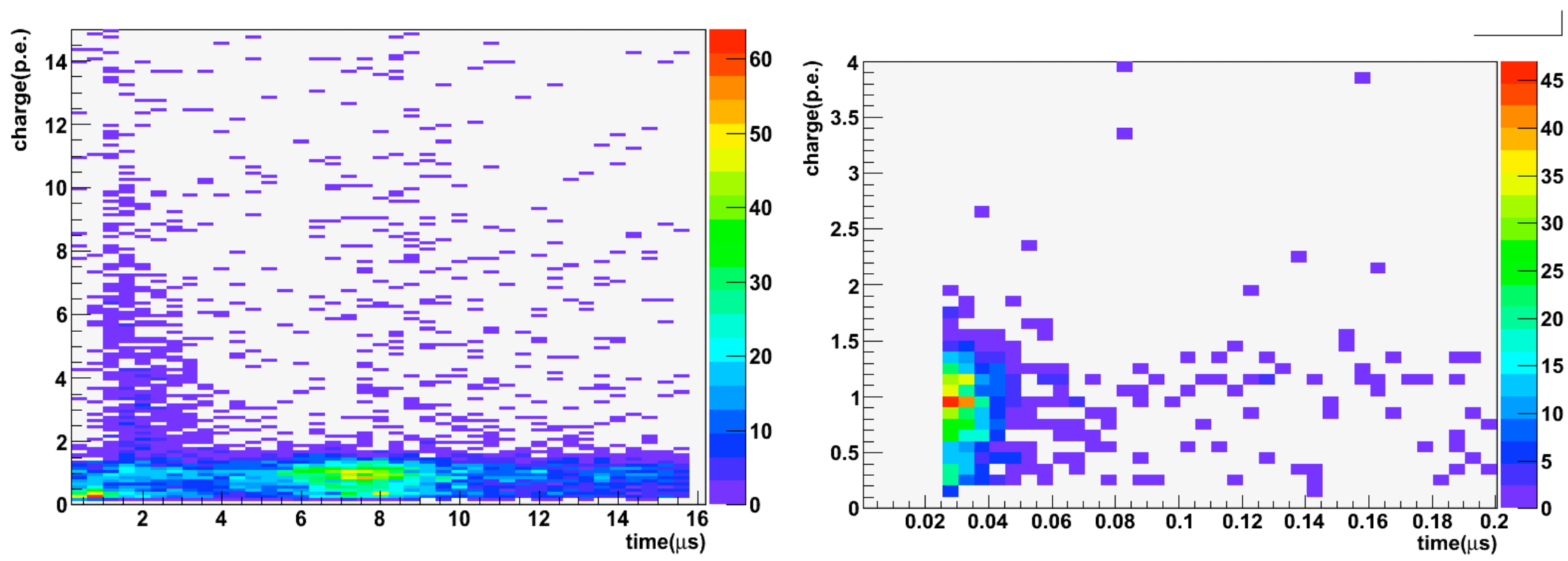

***FIG.*** *7 – The time vs amplitude distribution of after-pulses type 1a&1b (left) or type 2(right).*

In this case the events occur in the 25-40 ns interval after the main pulse as expected by the origin of these pulses while the amplitude is one pe on average.

The right plot represents the time vs. amplitude distribution of the type 2 after-pulses in the 200 ns-16 μs time interval after the main pulse. In this case more structures are observed with respect to the previous case. The large number of after-pulses occurring at 1-2 μs and 7-8μs after the main pulse corresponds to the scattering of the photo electron on light or heavy ions. It is important to observe that a sizeable fraction of these pulses have large amplitudes in accordance to the description of par. 1. In addition to this, we observe a region with small charge (≈ 1/3 p.e.) in the 500-1000 ns time interval after the main pulse. A zoom of this region is shown in Fig. 8a) while an event of this type as seen from the CAEN digitizer is shown in Fig. 8b). These events look as real after pulses of very small amplitude. The small amplitude is in favor of an event occurring in the dynodes region but the time window immediately below the 1-2 μs expected from the drift of light ions does not clarify the real origin of this type of observed after pulse

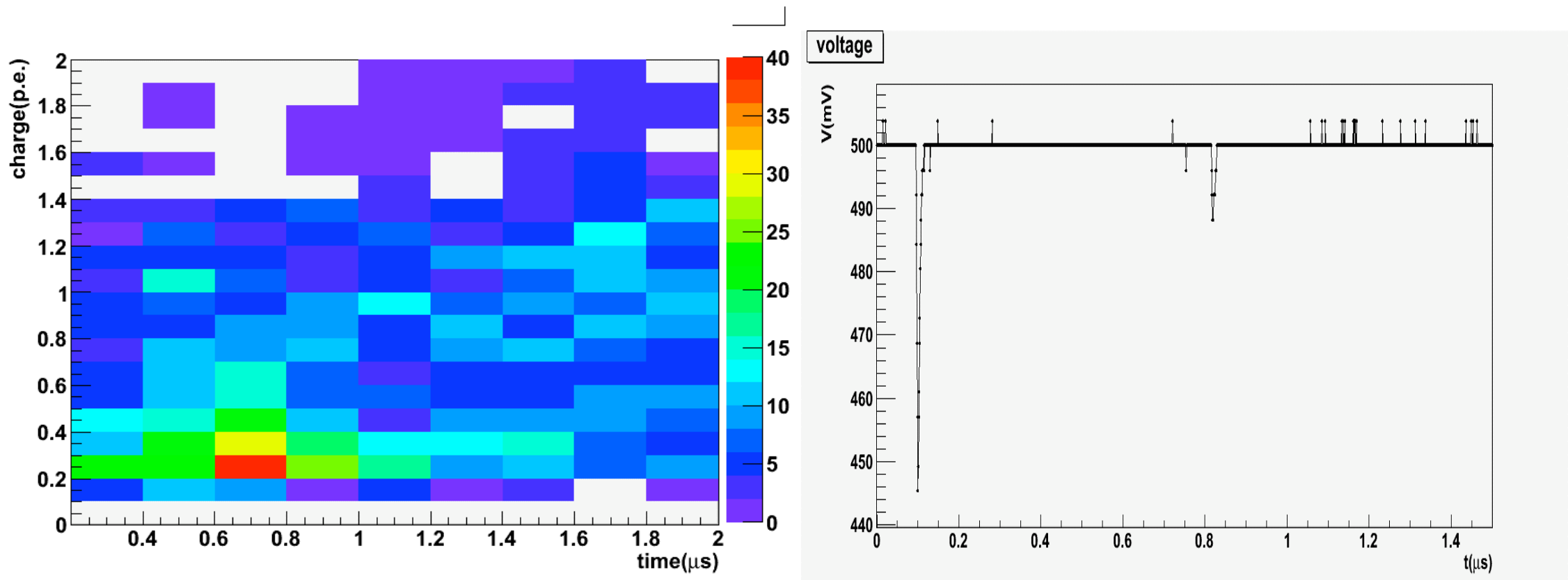

***FIG. 8*** *– a) zoom on the low-amplitude after-pulse region at around 0.5-0.8 μs after the main pulse (left); b)an example of these pulses is shown in the right (small signal occurring at 0.8 μs).*

In Fig. 9 the time distributions obtained from the plot in Fig.7 are shown. Most of the after-pulse of type 1a and 1b(left) occur after 25-40 ns the main pulse as expected from the design of the PMT and from the voltage supply. The two regions already observed for type 2 after-pulses at 2 and 8 μs are clearly evident though the structure of the heavy ion peak as described in [1] is not seen in our measurement. The plot includes the after-pulses which may occur more than once after the main pulse (sometimes 2 or even3 after-pulses are observed). The total contribution of these after-pulses is about 1.2%. and 8.1 % for type 1 and type 2 respectively.

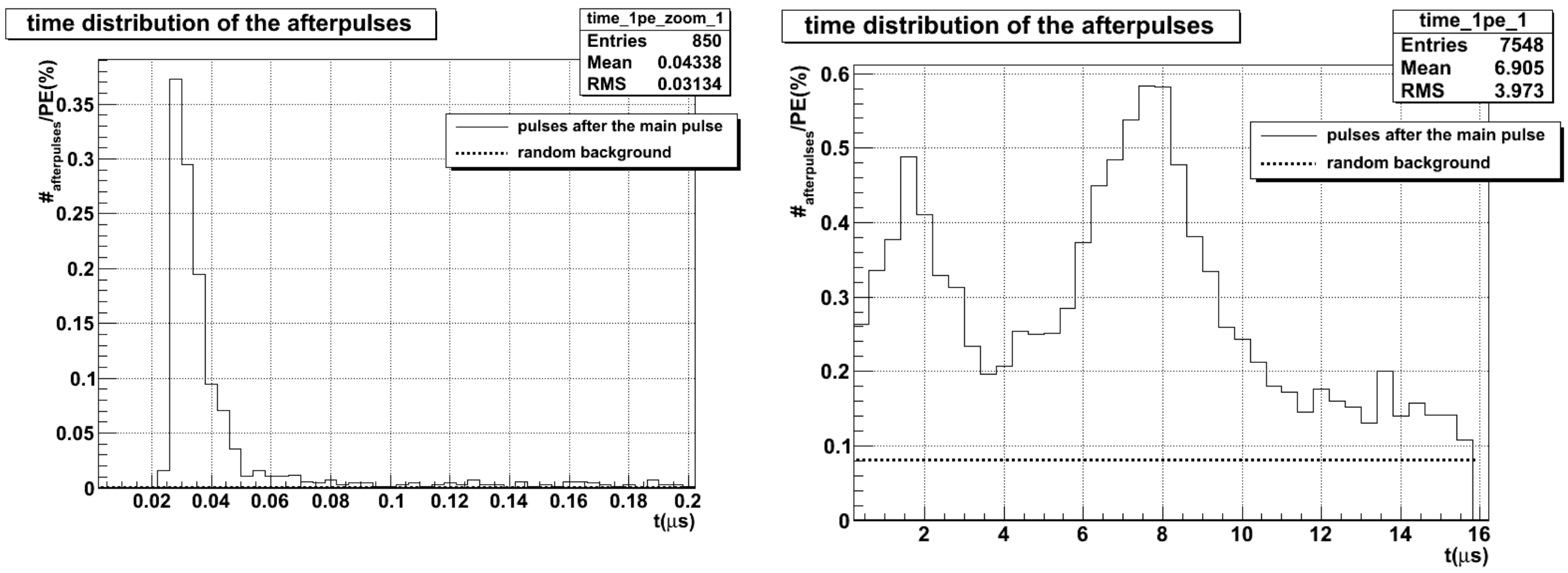

***FIG. 9*** *– Arrival time of the after-pulses after the main pulse normalized to the number of the main pulses in percents.*

**4 CONCLUSIONS**

The late- and after-pulses on the improved quantum efficiency, large area, Hamamatsu R7081MOD photomultiplier was measured with an experimental set up consisting of a light tight box, a laser light source and a 1GhZ CAEN digitizer to record the signals. The main features of these pulses were found consistent with respect to previous measurements on other large area PMTs[1,2] though a larger percentage of type 2 after-pulses is present in this high quantum efficiency Phototube. The time distribution of the late-pulses was carefully measured. The knowledge of this response function is particular important to a correct reconstruction of the muon tracks in underwater detectors where it may mimic the scattering of the Cherenkov photons in the water. The same applies to the observed large amplitude after pulses which could erroneously be ascribed to the intense light produced by muon tracks close to the PMT. The inclusion of the present measured effects in the MonteCarlo is expected to improve the agreement between simulations and experimental data.